\documentclass[%
 aip,
 floatfix,
 amsmath,amssymb,
 reprint,%
]{revtex4-1}

\usepackage{graphicx}
\usepackage{dcolumn}
\usepackage{bm}

\usepackage[utf8]{inputenc}
\usepackage[T1]{fontenc}
\usepackage{mathptmx}
\usepackage{etoolbox}
\usepackage[colorlinks, citecolor={blue}, urlcolor={blue}, linkcolor={blue}]{hyperref}

\usepackage{footmisc} 

\makeatletter
\def\@email#1#2{%
 \endgroup
 \patchcmd{\titleblock@produce}
  {\frontmatter@RRAPformat}
  {\frontmatter@RRAPformat{\produce@RRAP{*#1\href{mailto:#2}{#2}}}\frontmatter@RRAPformat}
  {}{}
}%
\makeatother

\begin{document}

\preprint{AIP/123-QED}

\title{Design and Characterization of Compact Acousto-Optic-Deflector Individual Addressing System for Trapped-Ion Quantum Computing}

\author{Jiyong Yu}
\affiliation{Duke Quantum Center, Duke University, Durham, NC 27701, USA}
\affiliation{Department of Electrical and Computer Engineering, Duke University, Durham, NC 27708, USA}
\author{Kavyashree Ranawat}
\affiliation{Duke Quantum Center, Duke University, Durham, NC 27701, USA}
\affiliation{Department of Electrical and Computer Engineering, Duke University, Durham, NC 27708, USA}
\author{Andrew Van Horn}
\affiliation{Duke Quantum Center, Duke University, Durham, NC 27701, USA}
\affiliation{Department of Electrical and Computer Engineering, Duke University, Durham, NC 27708, USA}
\author{Jacob Whitlow}
\altaffiliation[Current address: ]{IonQ, Inc., College Park, MD 20740, USA}
\affiliation{Duke Quantum Center, Duke University, Durham, NC 27701, USA}
\affiliation{Department of Electrical and Computer Engineering, Duke University, Durham, NC 27708, USA}
\author{Seunghyun Baek}
\affiliation{Department of Nano Science and Technology \& SKKU Advanced Institute of Nanotechnology (SAINT), Sungkyunkwan University, Suwon, 16419, Korea}
\author{Junki Kim}
\affiliation{Department of Nano Science and Technology \& SKKU Advanced Institute of Nanotechnology (SAINT), Sungkyunkwan University, Suwon, 16419, Korea}
\affiliation{Department of Quantum Information Engineering, Sungkyunkwan University, Suwon 16419, Korea}
\author{Jungsang Kim}
\email{jungsang@duke.edu}
\affiliation{Duke Quantum Center, Duke University, Durham, NC 27701, USA}
\affiliation{Department of Electrical and Computer Engineering, Duke University, Durham, NC 27708, USA}

\date{\today}

\begin{abstract}
We present a compact design for a beam-steering system based on acousto-optic-deflectors (AODs) used as an individual addressing system for trapped-ion quantum computing. The design targets to minimize the optomechanical degrees of freedom and the optical beam paths to improve optical stability, and we successfully implemented a solution with a compact footprint of less than 1 square foot. The system characterization results show that we achieve clean Gaussian beams at 355nm wavelength with a beam steering range of $\sim$50 times the beam diameter, and an intensity crosstalk of $< 9 \times 10^{-4}$ at all neighboring ions in a five-ion chain. Based on these capabilities, we experimentally demonstrate individual addressing of a 30-ion chain. We estimate the beam switching time of the AOD to be $\sim$240 ns. The compact system design is expected to provide high optical stability, providing the potential for high-fidelity trapped-ion quantum computing with long ion chains.
\end{abstract}


\maketitle

\section{Introduction}
\label{sec:introduction}
Quantum computing can provide solutions to problems that are intractable for classical computers \cite{shor1999polynomial}, and recent experiments have demonstrated the potential for quantum advantages \cite{arute2019quantum, madsen2022quantum}. The trapped-ion system is a particularly promising platform
due to high-quality qubits and high-fidelity quantum gate operations \cite{bruzewicz2019trapped}. Recent experiments have shown near-perfect state preparation and measurement (SPAM) \cite{myerson2008high, noek2013high, crain2019high, christensen2020high, sotirova2024high}, very long coherence time exceeding one hour \cite{wang2017single, wang2021single}, and high-fidelity entangling gate operations \cite{ballance2016high, gaebler2016high, wang2020high, clark2021high, baldwin2021high, loschnauer2024scalable, hughes2025trapped}, primarily through ion-laser or ion-microwave interactions. 

Precise individual addressing of each ion qubit in a long chain is also a crucial ingredient for a scalable trapped-ion quantum computing architecture \cite{zhu2006trapped, lin2009large, monroe2013scaling}. This architecture provides a path to implementing generic algorithms on quantum processors with all-to-all connectivity, without the time-consuming shuttling and cooling operations between gates \cite{moses2023race, loschnauer2024scalable}. Several implementations of laser individual addressing systems have been demonstrated, including micro-electromechanical systems (MEMS) \cite{wang2020high, crain2014individual}, multi-channel acousto-optic modulators (AOMs) \cite{debnath2016demonstration, wright2019benchmarking, kim2020hardware}, acousto-optic deflectors (AODs) \cite{pogorelov2021compact, Chen2024benchmarkingtrapped, chen2024low}, and integrated photonic chips \cite{sotirova2024low}. Among these methods, AODs can tolerate a higher optical power budget at UV wavelength than MEMS mirrors, allowing for faster quantum gate operations. In addition, the continuous steerability of AODs alleviates the even ion spacing condition for a long ion chain, which is typically required for multi-channel AOMs or integrated photonic chips.

In this work, we present the design and characterization of a compact AOD-based individual addressing system for scalable trapped-ion quantum computing. We utilized innovative opto-mechanical designs to assemble a stable optical system with minimal degrees of freedom. We performed a full system characterization, including beam profile, beam steering range, intensity crosstalk, and beam switching time. The characterization results
aligns well with the system design target, demonstrating applicability to trapped-ion quantum computing with a long ion chain.

\section{System Design and Assembly}
\label{System Design and Assembly}
\subsection{Optical system design}
\label{Optical system design}

\begin{figure*}[t]
\centering
\includegraphics[width=\textwidth]{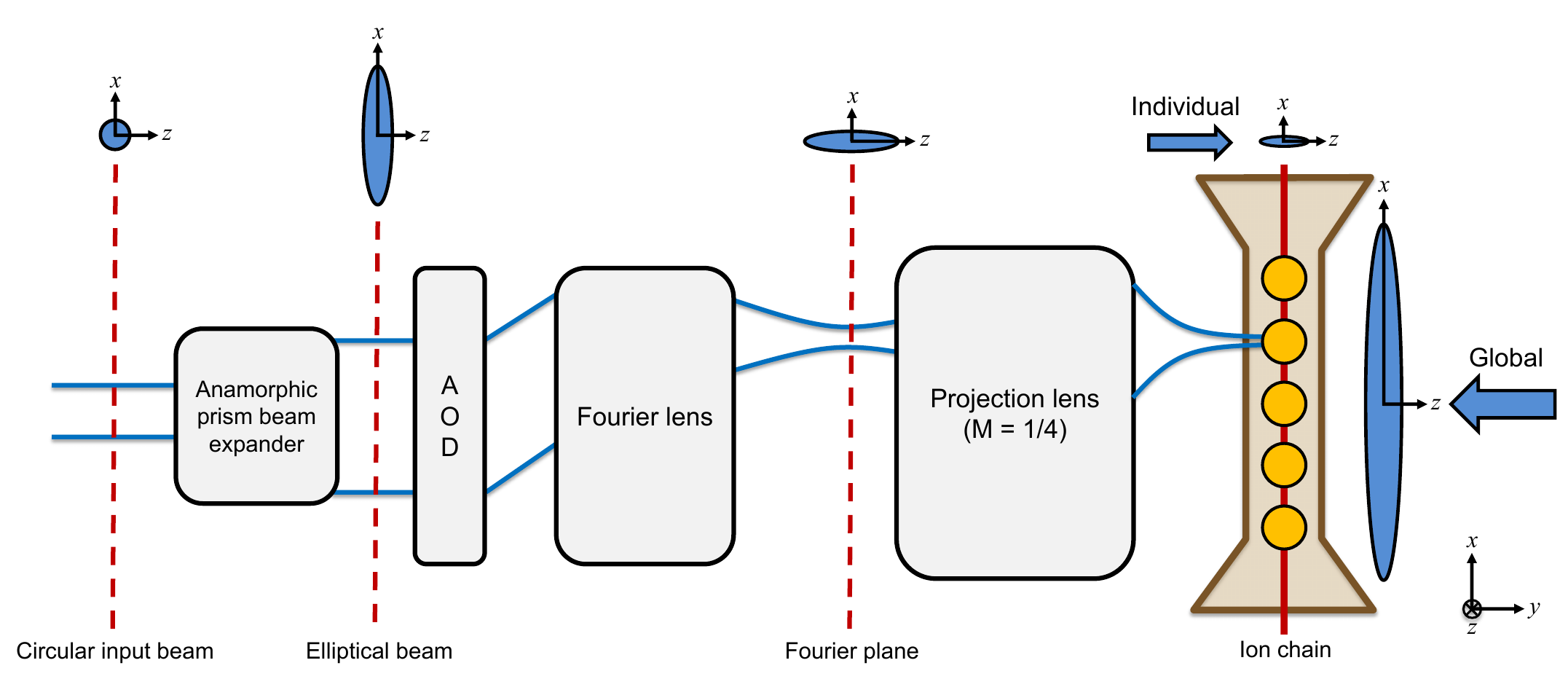}
\caption{Optical design schematic of the compact AOD individual addressing system. A custom-designed anamorphic prism beam expander
expands the circular collimated input beam with an aspect ratio of 4.7 along the $x$-axis, generating an elliptical beam with a beam waist of 0.32 mm by 1.5 mm. The generated elliptical beam passes through the AOD
(Brimrose CQD-150-100-355), which deflects the beam with a maximum deflection angle of 6 mrad over an RF
bandwidth of 100 MHz. A Fourier lens with a $\sim100$ mm
focal length is positioned $\sim100$ mm away from the AOD, converting
angular deflection from the AOD into parallel positional translation at
the Fourier plane. The semi-focused beam on the Fourier
plane has a target design waist of 34 $\mu$m x 7.3 $\mu$m and
a steering range of 600 $\mu$m. The final focusing at the ion
position is achieved with an NA=0.2 projection lens (Photon Gear 18020-ATP), providing a 4X demagnification with
minimized optical aberration. The target beam waist at
the ion position is around 8.5 $\mu$m x 1.8 $\mu$m, with a maximum steering range of 150 $\mu$m.}
\label{fig1}
\end{figure*}

The compact AOD individual addressing system is designed to be integrated with our existing cryogenic trapped-ion quantum computing system 
\cite{spivey2021high}, in which we trap a chain of $^{171}\text{Yb}^+$ ions. The detailed description of the trapped ion system is provided in Ref.~\onlinecite{spivey2021high}. The hyperfine electronic ground states of the \(^{171}\text{Yb}^+\) ions are utilized as qubits to store quantum information \cite{olmschenk2007manipulation}. The qubit state can be manipulated through a two-photon Raman transition using a 355 nm pulsed laser (Coherent Paladin 355) \cite{debnath2016demonstration}. The 355 nm beam from the laser source is divided into two optical paths: one global beam that simultaneously illuminates all ions in a chain and two individual beams that selectively address each ion \cite{spivey2022compact}. The global beam and the individual beams are set up in a counter-propagating geometry, allowing the net momentum kick to couple with the radial motional modes of the ion chain.

Figure~\ref{fig1} illustrates the schematic optical design of the compact AOD individual addressing system. In our experiment, we utilize a microfabricated surface trap (Peregrine trap from Sandia National Laboratories) \cite{revelle2020phoenix} where the Raman beams propagate along the surface of the trap. We designed the final beam shape to be elliptical elongated perpendicular to the trap surface, to reduce the numerical aperture of the beam so that clipping at the trap edge will be minimized, while allowing tight focusing along the direction of the ion chain \cite{spivey2022compact}. A custom-designed anamorphic prism pair beam expander generates an elliptical beam with a precise aspect ratio. The elongated beam is deflected by an AOD and refocused with a Fourier lens, providing parallel beam steering capability at the Fourier plane. The final projection lens images the intermediate focused beam at the Fourier plane onto the ion position with a demagnification factor of 4. The target beam waists and steering range at the ion position are confirmed by simulation to be approximately 8.5 $\mu$m x 1.8 $\mu$m and 150 $\mu$m, respectively. Assuming the average ion spacing of $\sim5\mu$m, this provides an individual addressing capability of up to $\sim 30$ ions.

\begin{figure*}[t!]
\centering
\includegraphics[width=0.9\textwidth]{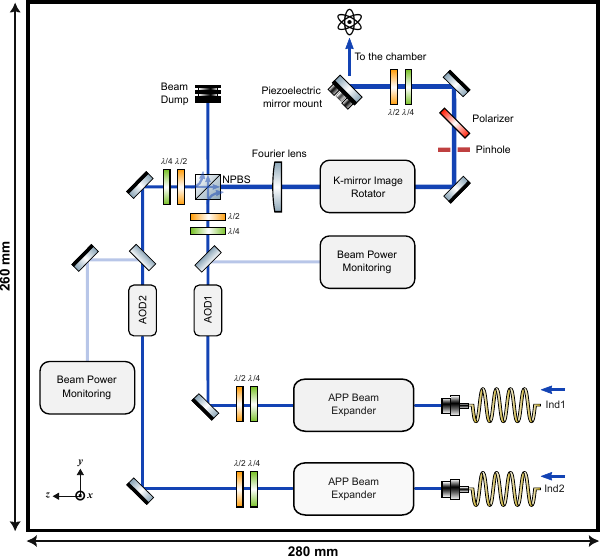}
\caption{
A detailed schematic diagram of the compact AOD individual addressing system. Two individual beams (Ind1 and Ind2) are routed from the upstream optical setup via photonic crystal fibers. After collimation, a compact anamorphic prism pair (APP) beam expander converts each individual beam into an elliptical beam. Each beam is deflected by its corresponding AOD along the $x$ axis. A beam sampler picks off $\sim\!0.2$\% of the deflected beam from AOD and directs it to a photodiode to monitor beam power fluctuations. A 50:50 non-polarizing beam splitter (NPBS) combines both individual beams spatially. A Fourier lens focuses the beams and transforms the angular deflection from AOD into parallel displacements at the Fourier plane. A K-mirror image rotator enables alignment of the beam steering direction along the ion chain. A precision optical pinhole is positioned at the Fourier plane to block 0th and 2nd-order diffraction from AODs. A linear P-polarizer filters the polarization of the final beam. Lastly, a piezoelectric kinematic mirror mount (Polaris-K05P2) provides fine adjustment of beam alignment at the ion position. The system features a compact footprint of less than 1 square foot.
}
\label{fig2}
\end{figure*}

\subsection{Compact optomechanical system assembly}
\label{Compact optomechanical system assembly}
Traditional atomic, molecular, optical (AMO) experimental systems employ conventional optical assemblies that utilize bulky opto-mechanical mounts on standard optical breadboards. While this approach offers system design flexibility, it generally lacks mechanical stability for high-fidelity quantum gate operations. Therefore, we adopted a compact system design approach with minimal control degrees of freedom and reduced optical beam path in order to enhance overall optical stability. Our approach incorporates several custom-designed optomechanical components and precise alignment protocols to achieve this goal.

The layout of the compact optomechanical system assembly is summarized in Figure~\ref{fig2}. Two individual beams are routed to the system from upstream optical module \cite{spivey2022compact} via photonic crystal fibers (PCF) \cite{colombe2014single}. The upstream optical module provides frequency, phase, and amplitude modulation of the Raman beams, including compensation of the frequency shift caused by the AOD as the beam is scanned across the ion chain. The compact anamorphic prism pair (APP) transforms each circular collimated beam into an elliptical beam. Each beam then passes through an AOD mounted on a custom-designed flexure assembly. A fraction of the deflected beam is sampled using a beam sampler and focused onto a photodiode to monitor optical power fluctuations.

The two individual beams are then spatially overlapped using a non-polarizing beam splitter (NPBS) and projected through a Fourier lens and a K-mirror image rotator. A K-mirror image rotator, consisting of three mirrors arranged in a ‘K’ configuration, rotates the image around the optical axis without introducing lateral displacement, and is widely used to correct image orientation \cite{10.1117/12.2312346}. The K-mirror image rotator is employed in our design to precisely align the beam steering direction of AOD along the length of the ion chain. A precision optical pinhole is positioned at the Fourier plane to block unwanted 0th and 2nd-order diffraction from the AOD. The final polarization of the beam is filtered using a linear P-polarizer with a high extinction ratio, which is necessary to minimize the four-photon differential Stark shift of the Raman transition \cite{debnath2016programmable}.

The system assembly footprint is less than 1 square foot with a total effective optical beam path length of $\sim\!51$ cm. We expect that this compact system volume with reduced optical beam path length provides enhanced optical stability, which is essential for high-fidelity quantum gate operations. We will provide detailed descriptions for each part of the system in the following sections.

\begin{figure}[t]
\centering
\includegraphics[trim={1cm 2.5cm 0.3cm 0.5cm}, clip, width=0.8\columnwidth]{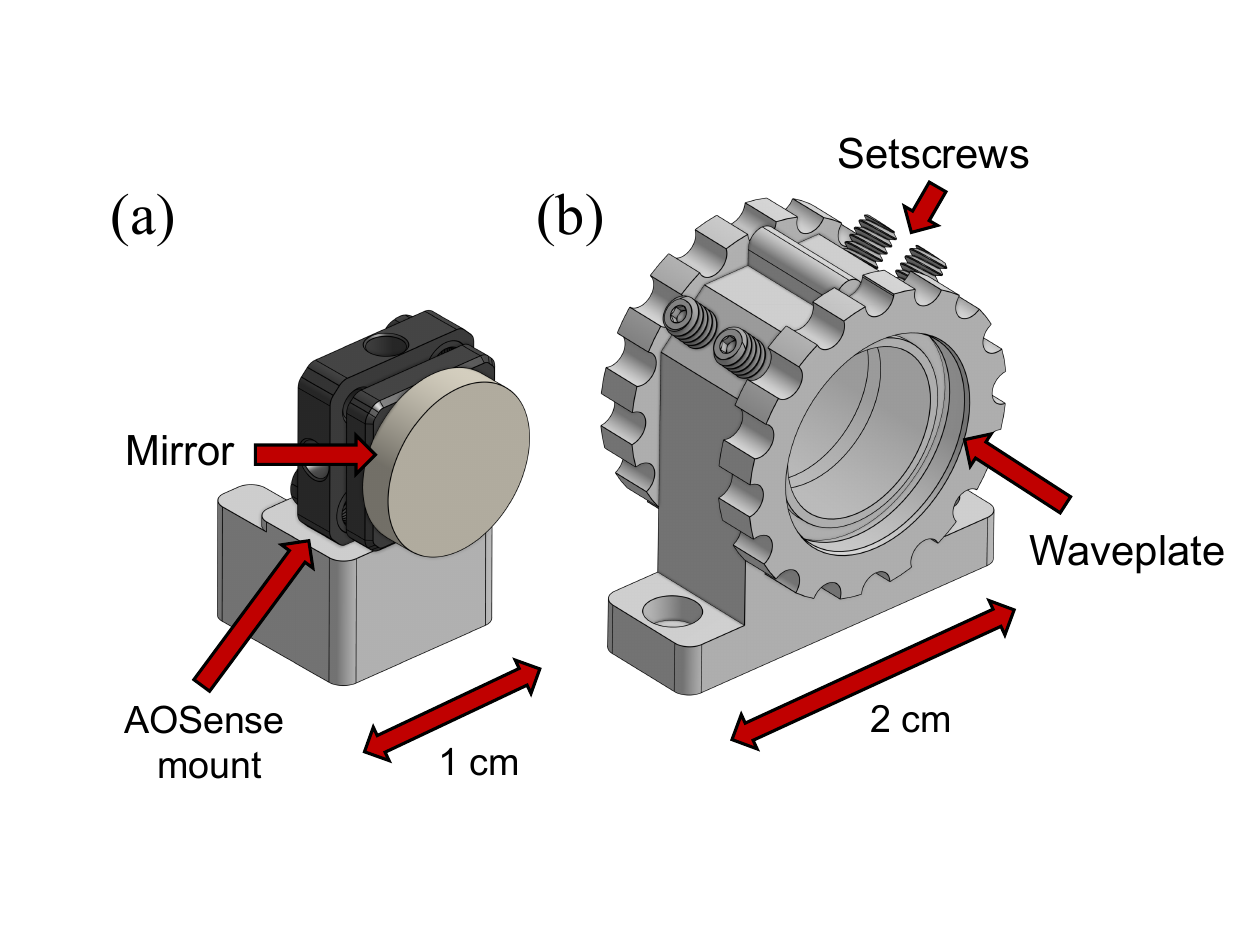}
\caption{
Compact optomechanical components for mounting mirrors and waveplates. (a) Mirrors are mounted on commercially available miniature kinematic mounts (AOSense). (b) Optical waveplates are mounted on custom-designed rotational mounts. A half-wave and quarter-wave plate can be mounted in pairs to provide arbitrary polarization control. Setscrews provide mechanical locking for the waveplates.
}
\label{fig3}
\end{figure}

\begin{figure}[t]
\centering
\includegraphics[trim={0.5cm 1.5cm 1.0cm 0.5cm}, width=0.85\columnwidth]{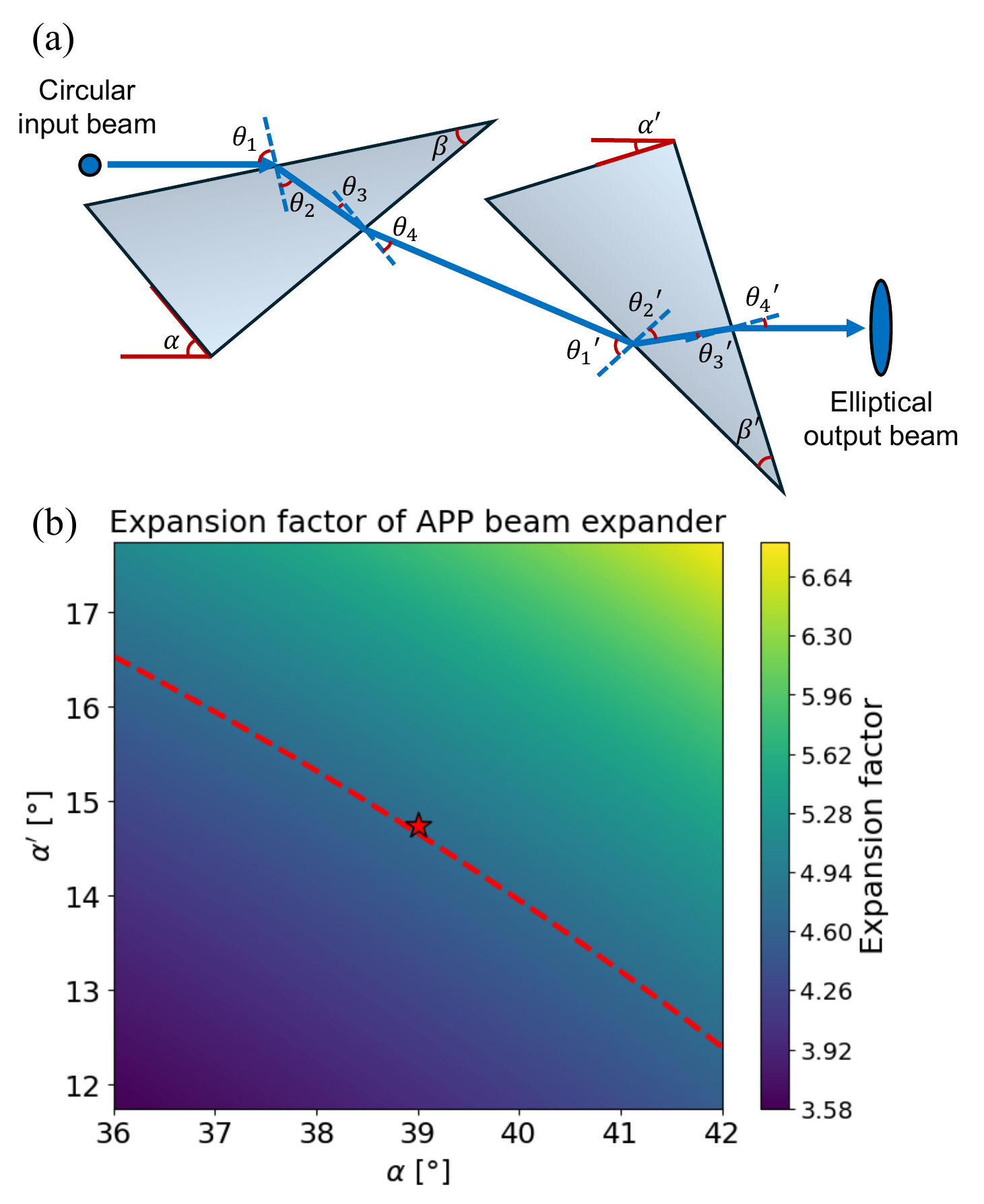}
\caption{
(a) Schematic of the anamorphic prism pair (APP) beam expander with all geometric angles defined. $\alpha, \alpha'$ and $\beta, \beta'$ are major factors for determining the beam expansion factor. (b) Numerical simulation of the APP beam expander beam expansion factor. We used $\beta=\beta'$ of 30$^{\circ}$ in the simulation. The red dotted line represents the condition satisfying the target expansion factor of 4.7. The red star represents the actual system design parameters ($\alpha$, $\alpha'$) = (39.0$^\circ$, 14.75$^\circ$).
}
\label{fig4}
\end{figure}

\subsection{Compact optomechanical components}
\label{Compact optomechanical components}
We used miniature optomechanical components to replace conventional bulky parts. Figure~\ref{fig3}(a) shows a commercially available kinematic mount (AOSense) used for mirrors, and Figure~\ref{fig3}(b) shows a custom-designed miniature rotational mount for optical waveplates. These components have a footprint of less than 2 cm², enabling a compact and efficient beam arrangement on the base plate. Once the optical alignment is done, all component positions can be locked mechanically, providing long-term optical stability.

\subsection{Compact anamorphic prism pair beam expander}
\label{Compact anamorphic prism pair beam expander}

\begin{figure}[t]
\centering
\includegraphics[width=\columnwidth]{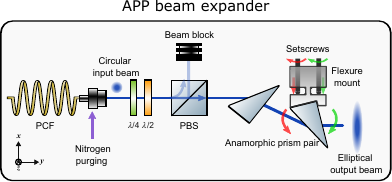}
\caption{A simplified schematic diagram of the compact anamorphic prism pair (APP) beam expander is presented. Non-polarization-maintaining PCF (NKT Photonics LMA-10-UV-FUD) routes the beam from the upstream into a nitrogen-purgeable collimator (Schäfter + Kirchhoff 60FC-4-S18-49-XV). Nitrogen purging can prevent fiber tip degradation induced by particle deposition at UV wavelengths \cite{marciniak2017towards}. The collimator is mounted on a kinematic mount for the initial alignment process. A polarizing beam splitter (PBS), together with a quarter-wave plate and half-wave plate, provides initial polarization filtering. The geometry of PBS is set properly so that the transmission of the polarization along the $x$ axis is maximized, at which the anti-reflection coating of the anamorphic prism surfaces is designed to be optimal. The second anamorphic prism is mounted on a flexure mount, allowing for fine adjustment of the relative angles between the prisms. Each setscrew provides clockwise or counterclockwise rotation of the second prism.
}
\label{fig5}
\end{figure}

As described in Section~\ref{Optical system design}, the circular output beam from the fiber collimator is converted to an elliptical beam with a desired aspect ratio. A cylindrical lens pair can provide this feature, but finding an off-the-shelf lens combination that accurately matches the target aspect ratio can be challenging. An alternative solution is an anamorphic prism pair (APP) beam expander \cite{kasuya1978prism}, which provides flexible control of the expansion factor by adjusting the relative angle between the prisms. 

Figure~\ref{fig4}(a) illustrates the typical geometry of APP that generates an elliptical beam. The relationships between each angle and the total beam expansion factor can be calculated analytically using simple ray optics \cite{kasuya1978prism}. We conducted a numerical simulation using the analytical formula to determine the design parameters for our APP beam expander. We used fused-silica anamorphic prisms with a refractive index of $ n\sim 1.476$ at 355 nm and wedge angle $\beta=30.0^{\circ}$. Figure~\ref{fig4}(b) shows the contour plot of the beam expansion factor with the target condition represented as a red-dotted line. We chose the actual system design parameter ($\alpha$, $\alpha'$)
= ($39.0^{\circ}$, $14.75^{\circ}$), which is close to the target expansion factor of 4.7.

However, the beam expansion factor of an actual APP beam expander can deviate significantly from the target value. The primary source of error is the finite tolerance of the machined parts and the prisms. The tolerances of the major angles in Figure~\ref{fig4}(a) are $\lvert\Delta\alpha\rvert=1.0^{\circ}$, $\lvert\Delta\alpha'\rvert=1.0^{\circ}$, $\lvert\Delta\beta\rvert=0.25^{\circ}$, and $\lvert\Delta\beta'\rvert=0.25^{\circ}$. These tolerances correspond to beam expansion factor errors of approximately $7\%$, $5\%$ $2\%$, and $1\%$, respectively. Additional errors can occur during the assembly process, which is not trivial to measure precisely. As a result, the worst-case error of $\pm15\%$ or more can be generated for the beam expansion factor. The actual beam size measured during the assembly process was $\sim\!10\%$ smaller than the ideal value, which is consistent with the tolerance budget. 

We compensated for this error by mounting one of the prisms on a bi-directional flexure mount, as described in Figure~\ref{fig5}. This setup enables fine-tuning of the angle $\alpha'$ by rotating the second prism in either direction. We compensated for the measured beam expansion factor error with two setscrews, achieving an exact beam expansion factor of 4.7. 

\subsection{Flexure assembly for AOD alignment}
\label{Flexure assembly for AOD alignment}

\begin{figure}[t]
\centering
\includegraphics[width=0.85\columnwidth]{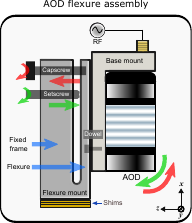}
\caption{A simplified schematic for flexure-mounted AOD is illustrated. The AOD is fixed on the base mount, which is attached to the flexure mount using a dowel pin and two capscrews. The dowel pin serves as a pivot point, allowing the base mount to rotate against the flexure mount before being fastened. This enables precise Bragg angle matching to achieve the optimal diffraction efficiency of the AOD. A capscrew or setscrew is installed on the flexure mount. Fastening the capscrew (setscrew) pulls (pushes) the base mount, rotating the AOD in a counterclockwise (clockwise) direction in the figure. Stainless steel shim stocks with a thickness of $\sim$250 $\mu$m are used to compensate for the positional shift of the AOD caused by the rotation.
}
\label{fig6}
\end{figure}

\begin{figure}[t]
\centering
\includegraphics[width=1.0\columnwidth]{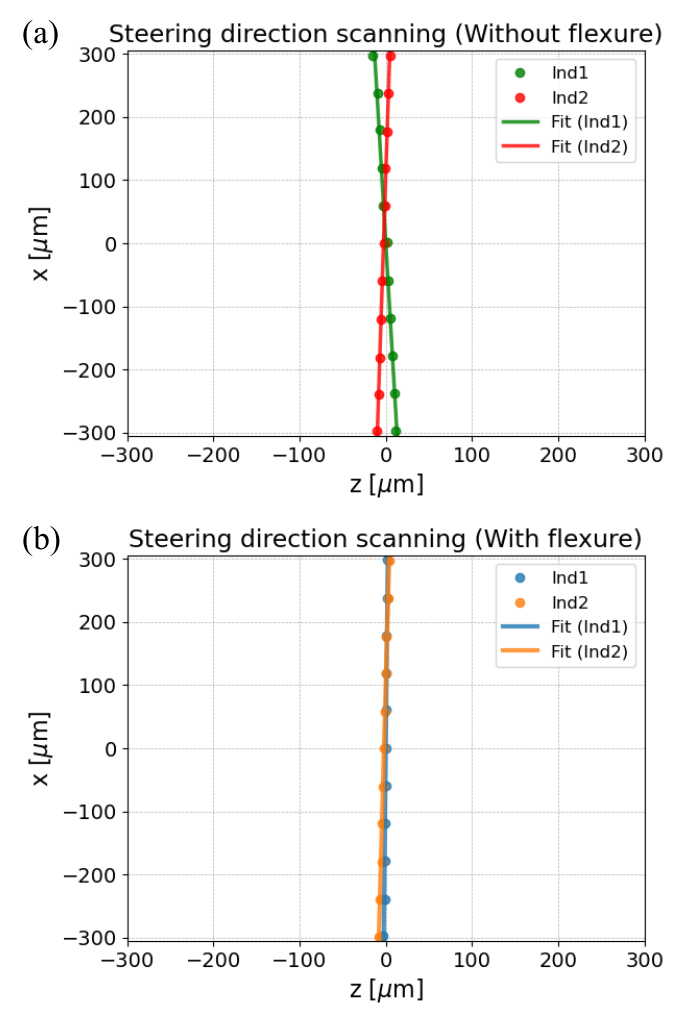}
\caption{
Beam steering direction measurement of two individual beams (Ind1 and Ind2) on the Fourier plane. The measured beam steering range is 600 $\mu$m. (a) The steering direction alignment error is $\sim$3.8$\boldsymbol{^\circ}$ when AODs are mounted on non-flexure mounts. (b) The steering direction alignment error is reduced to $\sim$0.7$\boldsymbol{^\circ}$ for flexure-mounted AODs.
}
\label{fig7}
\end{figure}

The steering directions of both individual beams must overlap precisely to ensure simultaneous addressing capability across the ion chain. However, manufacturing tolerances of AODs can lead to significant misalignment of their steering directions if simply referenced to the AOD housings. These misalignments can result in the shift of the Gaussian beam center away from the ions as the beam is steered across the chain. For example, for a 30 ion chain with an average ion spacing of $\sim5\mu m$, steering direction misalignment should be less than $1^{\circ}$ to achieve a power imbalance of less than $\sim\!\!10\%$ at the edge ion position. 

We utilized custom-designed flexure mounts that provide each AOD with the necessary degrees of freedom for precision alignment (Figure~\ref{fig6}). The flexure mount has a setscrew-tapped hole and a through-hole in the fixed frame, allowing us to both push and pull the flexure against the frame in a controllable manner. By mounting the AOD on the flexure, this design enables precise alignment of the acoustic wave directions of two different AODs. We used flexure mounts for both AODs to maximize the compensation range. The calibration results show that one full rotation of each screw provides the angular rotation of approximately $\pm 1^{\circ}$. 

Figure~\ref{fig7} illustrates the compensation results achieved using the custom flexure mounts. We measured the steering direction of both individual beams on the Fourier plane by sweeping the frequency over the full bandwidth of the AOD (100-200 MHz). We measured the full steering range of 600 $\mu$m for both beams on the Fourier plane. The expected full steering range at the ion position is thus 150 $\mu$m, corresponding to 30 ions with an average target separation of 5 $\mu m$. In Figure~\ref{fig7}(a), the initial steering direction alignment error between the two AODs was $\sim3.8^{\circ}$. By adjusting the direction of two AOD deflections with the flexure mount, we were able to reduce the alignment error of the two deflection directions to below the target value of $1^{\circ}$. Figure~\ref{fig7}(b) shows the reduced alignment error of $\sim0.7^{\circ}$ using the flexure mounts.

However, the rotation of AOD using the flexure mounts can shift the AOD crystal position, causing a reduction in the AOD diffraction efficiency. We estimate that a rotation of 1$^\circ$ generates a positional shift of the AOD crystal center by a $\sim$300 $\mu$m in both horizontal and vertical directions. To compensate for this shift, we used stainless steel shim stocks with a $\sim$250 $\mu$m thickness. We achieved near-perfect AOD diffraction efficiency with a few rounds of trial and error when inserting shim stocks during the assembly process.

\subsection{K-mirror image rotator}
\label{K-mirror image rotator}

\begin{figure}[t]
\centering
\includegraphics[width=0.95\columnwidth]{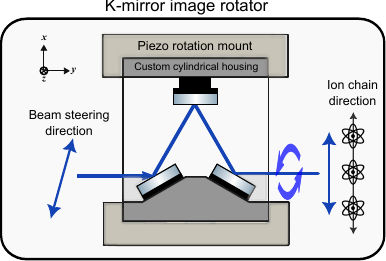}
\caption{A simplified section view of a K-mirror image rotator. Three mirrors are arranged in a triangular configuration. The center mirror is mounted on a kinematic mount (AOSense) to precisely align the optical axes of the input and output beams. The mirrors are mounted inside a custom cylindrical housing, and the housing is fixed on a piezo rotation stage (Newport AG-PR100), enabling precise alignment of the output beam steering direction along the ion chain direction.
}
\label{fig8}
\end{figure}

Even though the beam steering direction of the two AODs can be aligned precisely with each other using flexure mounts, the aligned steering direction can still deviate from the actual axial direction of the ion chain. A Dove prism \cite{fang2023improving} or K-mirror image rotator \cite{10.1117/12.2312346} can provide an additional rotational degree of freedom to compensate for this deviation. We utilized a K-mirror image rotator to avoid possible optical aberration induced by the Dove prism. Figure~\ref{fig8} represents a schematic of the K-mirror image rotator assembly. The assembly consists of three mirrors arranged in a triangular configuration and is mounted on a piezo rotation stage, which provides fine rotational adjustment. This assembly allows a precise alignment of the beam steering direction with the ion chain direction, which is essential for an experiment with a long ion chain.

\subsection{Beam power monitoring}
\label{Beam power monitoring}

\begin{figure}[t]
\centering
\includegraphics[width=1.0\columnwidth]{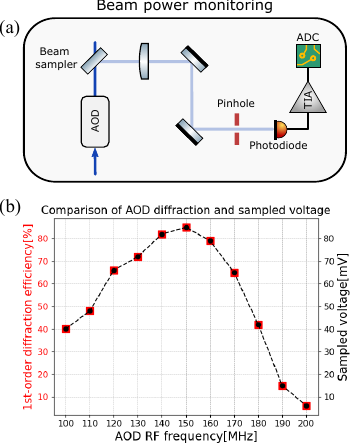}
\caption{
(a) A schematic diagram for the beam power monitoring system is illustrated. A beam sampler (Thorlabs BSF05-UV) samples $\sim0.2\%$ of the total beam power. A focusing lens with a focal length of $\sim75$ mm focuses the beam on the photodiode (Thorlabs SM05PD1A). A precision optical pinhole is installed before the photodiode to selectively monitor the 1st-order diffraction of the AOD. The current signal from the photodiode is converted into a voltage by a transimpedance amplifier (TIA) and subsequently sampled by an analog-to-digital converter (ADC) for continuous monitoring. (b) Comparison of the AOD diffraction efficiency and the sampled voltage after a TIA with a gain of 100 kV/A. The sampled beam power was $\sim$100 $\mu$W. The red square points represent the measured diffraction efficiency of the AOD, while the black circular points show the sampled voltage, showing a perfect overlap.
}
\label{fig9}
\end{figure}

\begin{figure*}[t]
\centering
\includegraphics[trim={2cm 0.5cm 1cm 2cm}, width=1.0\textwidth]{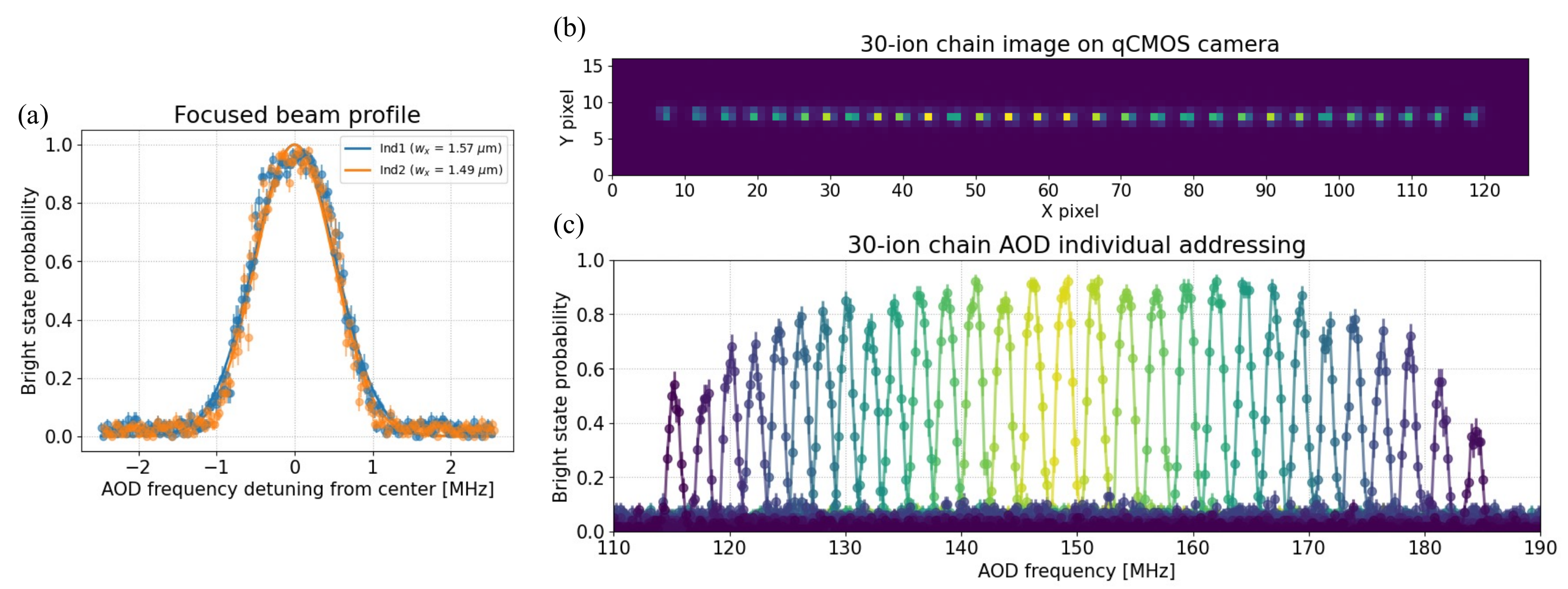}
\caption{
(a) In-situ measurement of beam profile for two individual beams (Ind1 and Ind2) using a single ion. The estimated beam waists are 1.57(1) $\mu$m for Ind1 and 1.49(1) $\mu$m for Ind2, respectively. (b) 30-ion chain image on the qCMOS camera with the imaging scale of 1 $\mu m$/pixel. The average spacing between the nearest ions is $\sim3.8$ $\mu m$. (c) Beam steering result of the AOD individual addressing system using 30 trapped ions. Well-resolved 30 Gaussian peaks represent the successful individual addressing along the entire chain. 
}
\label{fig10}
\end{figure*}

High-fidelity trapped-ion quantum gate operations require stable optical beam power driving the gate\cite{wu2018noise}. An active beam power monitoring with a proper feedback mechanism can be utilized to stabilize the beam power. Figure~\ref{fig9}(a) shows the beam power monitoring system. We sampled $\sim\!0.2\%$ of the beam power with a beam sampler and selectively monitored the 1st-order diffraction from the AOD using a lens–pinhole combination. The current signal from the photodiode was converted to a voltage by a transimpedance amplifier (TIA) and recorded as the AOD driving frequency was scanned. 
Figure~\ref{fig9}(b) shows the close agreement between the AOD diffraction efficiency and the sampled voltage, confirming reliable beam power monitoring. The converted voltage is sampled by an analog-to-digital converter (ADC) and can be employed in a digital feedback loop to stabilize the beam power \cite{zhang2021improving}.

\section{System characterization}
\label{System characterization}

\subsection{Beam profile}
\label{Beam profile}

We integrated the compact AOD individual addressing system into our current cryogenic trapped-ion system to measure the in-situ beam profile along the ion chain direction ($x$ axis). In our system, the fluorescence of each ion can be collected on different channels of a multi-mode fiber array, each connected to a photomultiplier tube (PMT), or on a single pixel of a qCMOS camera (Hamamatsu Photonics ORCA-Quest 2 qCMOS camera C15550-22UP)\cite{photonics2023qcmos}. The PMT-based collection method was used for qubit-state detection in all experiments, except for the 30-ion state detection in Section~\ref{Beam steering and individual addressing}.

Figure~\ref{fig10}(a) shows the results of the beam profiling measurement for each individual Raman beam (Ind1 and Ind2) using a single ion. We scanned the AOD RF frequency while driving a Rabi oscillation from an initially prepared dark state to the bright state, with the individual beam containing two RF tones. We measured the bright state probability at each point from which we can directly extract the beam waist. We measured the expected beam waist to be 1.57(1) $\mu m$ for Ind1 and 1.49(1) $\mu m$ for Ind2, respectively, using the expected steering efficiency of 1.5 $\mu$m/1 MHz. The estimated beam waists are slightly smaller than the target design value of 1.8$\mu m$, which might be attributed to the imperfect positioning of the final projection lens, which leads to an error in estimating the steering efficiency.

Along the axis perpendicular to the trap surface ($z$ axis), we conducted ex-situ through-the-focus measurement by installing a beam profiler near the expected virtual ion position. The measured beam waists along the $z$ axis were 9.7(1) $\mu m$ and 10.3(1) $\mu m$ for each individual beam, respectively. Although the measured beam waists are larger than the target design value of 8.5 $\mu m$, they still meet the clearance condition to avoid beam clipping from the trap edge \cite{spivey2022compact}.

\subsection{Beam steering and individual addressing}
\label{Beam steering and individual addressing}

We trapped a 30-ion chain to assess the in-situ beam steering and individual addressing performance. The average spacing between the nearest ions was set at approximately 3.8 $\mu$m by tuning the DC voltage solution. We utilized a qCMOS camera for the state detection of ions, since the number of the installed PMTs in our system was limited to 5. The fluorescence from each ion is collected by a 0.6 NA infinite-conjugate lens and designed to be focused onto a single pixel of the qCMOS camera. Figure~\ref{fig10}(b) represents the qCMOS image of 30 ions trapped with the DC voltage solution used in the experiment. Similar to the experiment described in Section~\ref{Beam profile}, we drove a Rabi oscillation using the individual beam while scanning the AOD RF frequency. Figure~\ref{fig10}(c) illustrates the beam steering results for the 30-ion chain, demonstrating the successful individual addressing performance along the entire 30-ion chain. 

\subsection{Intensity crosstalk}
\label{Intensity crosstalk}

\begin{figure}[t]
\centering
\includegraphics[width=0.81\columnwidth]{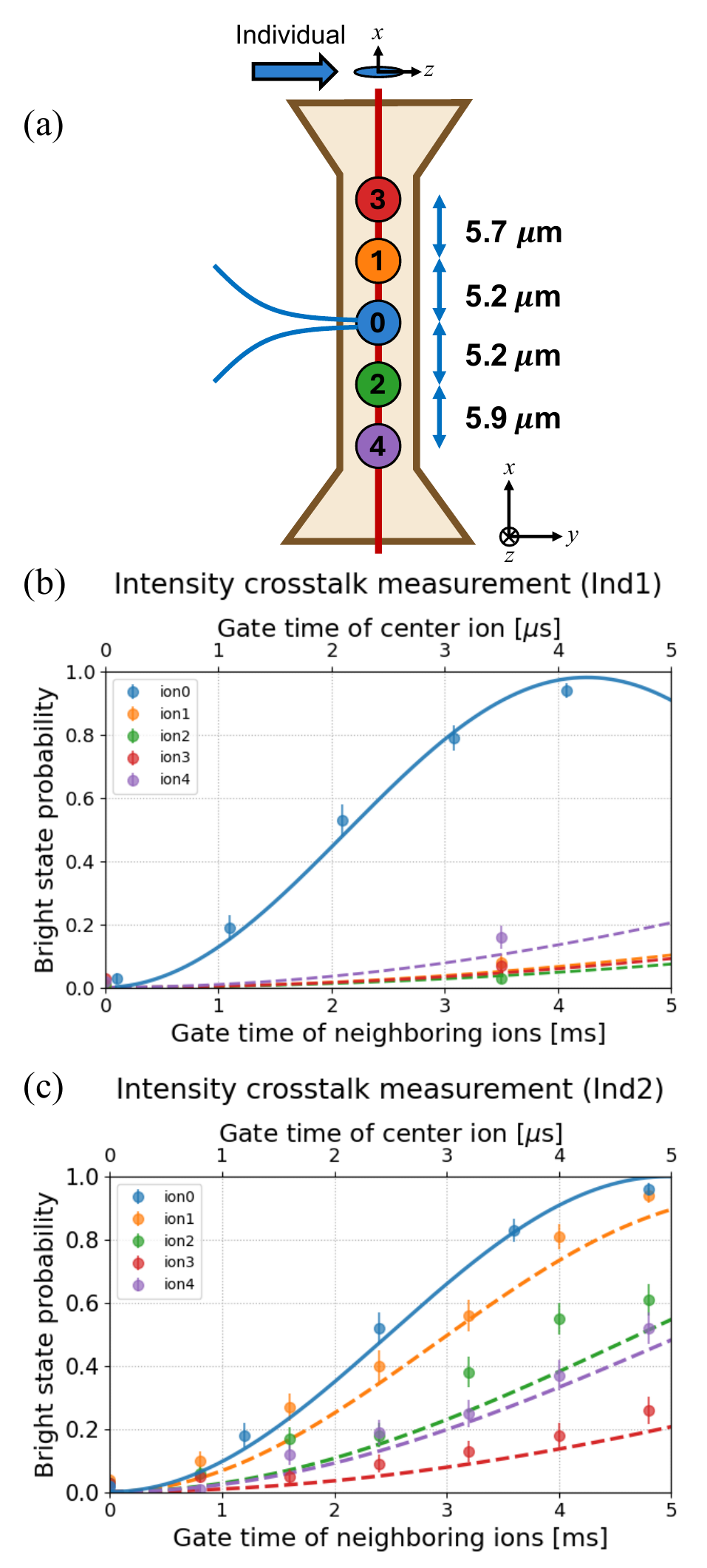}
\caption{
Intensity crosstalk measurement for each individual beam is presented. (a) Five ions are trapped, and the individual beam is aligned at the center ion (ion0). The distances between adjacent ions are represented. (b) First individual beam (Ind1) driving Rabi oscillations for all neighboring ions, $\pi\text{-time}$ ($t_{\pi}$) for ion0 is 4.26 $\mu s$. (c) Second individual beam (Ind2) driving Rabi oscillations for all neighboring ions, $t_{\pi}$ for ion0 is 4.98 $\mu s$. The gates were driven up to a maximum of 70 ms and 40 ms for Ind1 and Ind2, respectively, during which all neighboring ions showed coherent oscillations. For clarity, the maximum gate time in the figures for all neighboring ions is limited to 5 ms to emphasize the comparison with the center ion.
}
\label{fig11}
\end{figure}

The crosstalk between different qubits should be minimized for high-fidelity quantum gate operations \cite{wu2018noise}. In trapped-ion quantum computing gates driven by Raman beams, the primary source of this crosstalk arises from imperfect beam profiles induced by optical aberrations \cite{shih2021reprogrammable}. We measured the in-situ intensity crosstalk of our AOD individual addressing system using five trapped ions. We label ions in a chain as (3, 1, 0, 2, 4) as shown in Figure~\ref{fig11}(a). We aligned the individual beam at the center ion (ion 0) position, and drove Rabi oscillations using the individual Raman beam. Intensity crosstalk is estimated to be the ratio of the Rabi frequencies $\Omega_{neighbor}/\Omega_{center}$ for each neighboring ion. Figure~\ref{fig11}(b) and (c) show the intensity crosstalk measurement result for two different individual beams (Ind1, Ind2), respectively. The estimated intensity crosstalk at all neighboring ions is summarized in Table~\ref {tab:crosstalk}. The worst intensity crosstalk of Ind1 and Ind2 is estimated to be $2.6(1)\times10^{-4}$ and $8.6(3)\times10^{-4}$, respectively.

\renewcommand{\arraystretch}{1.3} 
\setlength{\tabcolsep}{12pt}
\begin{table}[h]
  \centering
  \caption{Estimated intensity crosstalk at neighboring ion positions for each individual beam.}
  \label{tab:crosstalk}
  \begin{tabular}{|c|c|c|} 
    \hline
    Ion label & Ind1 & Ind2 \\ 
    \hline
    1 & $1.8(1) \times 10^{-4}$ & $8.6(3) \times 10^{-4}$ \\
    2 & $1.6(1) \times 10^{-4}$ & $5.3(2) \times 10^{-4}$ \\
    3 & $1.7(1) \times 10^{-4}$ & $3.0(2) \times 10^{-4}$ \\
    4 & $2.6(1) \times 10^{-4}$ & $4.9(2) \times 10^{-4}$ \\
    
    \hline
  \end{tabular}
\end{table}

The ideal intensity crosstalk of a perfect Gaussian beam is expected to be $<1 \times 10^{-4}$ at the position of the neighboring ion, indicating that the measured intensity crosstalk is significantly higher than the ideal case. We expect that the primary sources of optical aberration are the compact anamorphic prism pair beam expander and the AOD, where the limited size of the optical apertures can cause undesired beam clipping. Various compensation techniques can be employed to actively cancel out the effect of the crosstalk during the quantum gate operations\cite{brown2004arbitrarily, fang2022crosstalk}.

\subsection{Beam switching time}
\label{Beam switching time}

\begin{figure}[t]
\centering
\includegraphics[width=1.0\columnwidth]{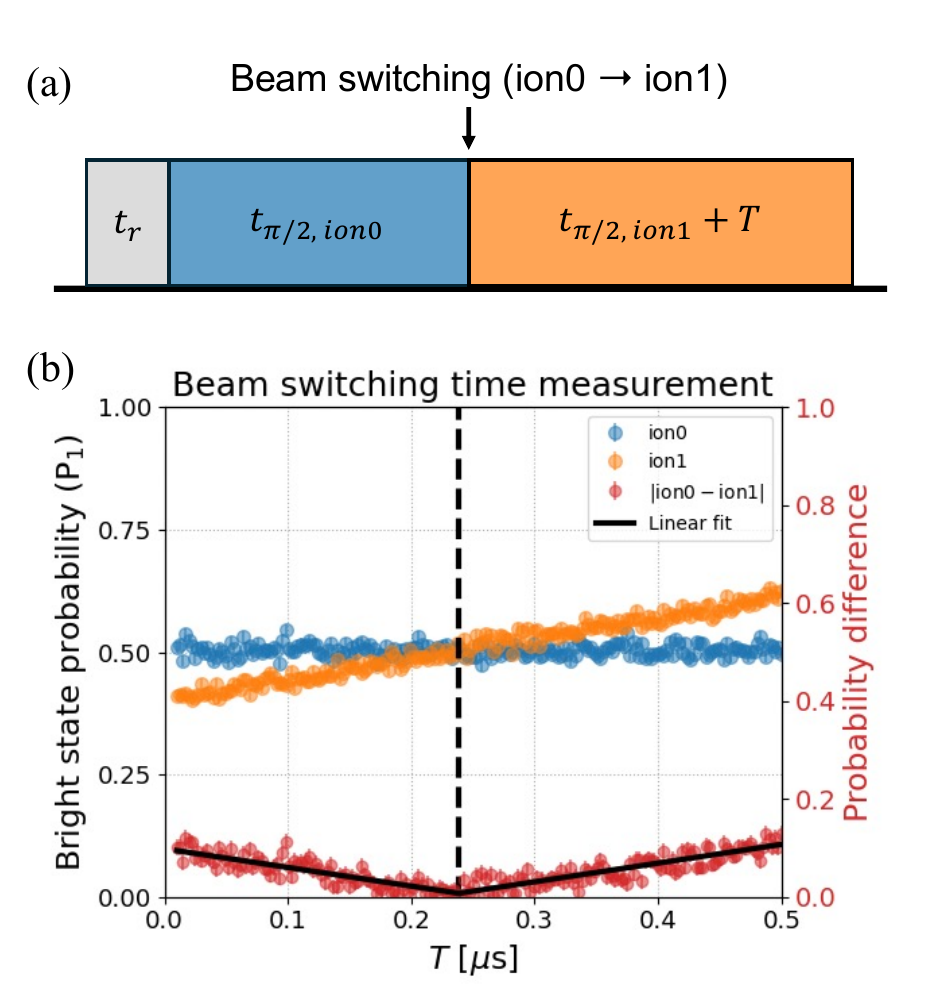}
\caption{
Beam switching time measurement experiment for two trapped ions with a separation of $\sim5 \mu m$. (a) A pulse sequence is used to measure the beam switching time; the individual Raman co-propagating gates drive each ion sequentially. Before the gate starts, RF signals are applied to both the AOD and the upstream AOM for a duration of AOM rising time $t_r\sim826$ ns. The $\pi/2$ pulse is applied to ion0, and the AOD RF frequency is switched to ion1 when the pulse is finished. Subsequently, $\pi/2$ pulse with additional driving time of $T$ is applied to ion1. We calibrated the $\pi/2$ time of each ion as 1.75 $\mu s$ and 1.74 $\mu s$, respectively. When the whole pulse sequence is finished, the bright state probabilities $\mathrm{P_{1}}$ of each ion are measured. The two-tone RF frequencies applied to the upstream AOM remain fixed throughout the entire gate sequence. (b) $\mathrm{P_{1}}$ of ion0 and ion1, and their absolute difference are plotted as a function of $T$. $\mathrm{P_{1}}$ of ion0 remain constant at 0.5 as expected, while $\mathrm{P_{1}}$ of ion1 increases as the additional driving time $T$ increases. The absolute difference of $\mathrm{P_{1}}$ between two ions is fitted (black curve), with a minimum at $T=238(3)$ ns (black dashed line), corresponding to the estimated beam switching time.
}
\label{fig12}
\end{figure}

We estimated the beam switching time $t_s$ of the AOD using two trapped ions separated by $\sim$5 $\mu m$. A Raman pulse scheme driving Rabi oscillation shown in Figure~\ref{fig12}(a) is used to measure the switching time with sufficient time resolution. First, we drive the first ion (ion0) for a $\pi/2$ time $t_{\pi/2,\ ion0}$. Right after the first pulse, we switch the beam to the second ion (ion1) by changing AOD frequency, and then drive ion1 for a duration of $(t_{\pi/2,\ ion1} + T)$. The bright state probabilities of both ions are measured when the whole sequence is completed.

Figure~\ref{fig12}(b) represents the bright state probability $\mathrm{P_{1}}$ of both ions as the time $T$ is varied. Ion0 shows the constant $\mathrm{P_{1}}$ of 0.5 as expected, while $\mathrm{P_{1}}$ of ion1 starts at less than 0.5, since the $\pi/2$ gate for ion1 is underdriven due to finite beam switching time. Therefore, we can expect that the $\mathrm{P_{1}}$ of ion1 will reach 0.5 at $T = t_s$, when the differences in $\mathrm{P_{1}}$ between ion0 and ion1 is minimized. Using this measurement scheme, we estimate the beam switching time $t_s$ of 238(3) ns. 

The expected beam switching time of AOD from the theoretical estimation is $1.3w_0/V=342$ ns\cite{1076309}, where the $w_0=1.5$ mm is the Gaussian beam waist at the AOD crystal and $V\sim5700$ m/s is the acoustic velocity inside the AOD crystal. The discrepancy from the experimental result may be attributed to the transducer geometry of the actual AOD device we used, as the theoretical estimation assumes an ideally flat rectangular shape for the AOD transducer \cite{1076309}. The measured beam switching time is short compared to typical trapped-ion quantum gate operation times and faster than the switching time of other individual addressing systems demonstrated, such as MEMS mirrors\cite{crain2014individual}.

\section{Conclusion}
\label{Conclusion}

In this work, we designed and characterized a compact AOD individual addressing system for trapped-ion quantum computing. Our compact approach of using miniature optomechanical components and assemblies allows for a system footprint of less than one square foot, potentially providing improved optical stability. We developed novel custom optomechanical assemblies, including an anamorphic prism pair beam expander, flexure mounts for AODs, and a K-mirror image rotator. These implementations provide an effective solution for achieving a precise target beam size and enable accurate alignment of the beam steering direction with that of a long ion chain.

The system design and characterization results demonstrate the potential applicability to scalable trapped-ion quantum computing or simulation with long ion chains. For example, a tightly focused individual beam with low intensity crosstalk on neighboring ions and a fast beam switching time can be utilized for mid-circuit qubit measurement and reset processes \cite{yu2025situ, chen2025non}. In addition, the combination of individual addressing capability with a broad beam steering range enables selective control of each individual ion in a long chain, which can be applied to the quantum simulation of the spin-boson model \cite{wang2024simulating, sun2025quantum} or bosonic systems \cite{katz2023programmable}.

\section*{Acknowledgements}
\label{Acknowledgements}

This work was primarily supported by the Office of the Director of National Intelligence - Intelligence
Advanced Research Projects Activity through ARO
contract W911NF-16-1-0082, the National Science Foundation STAQ Program PHY-1818914, the U.S. Department of Energy  Quantum Systems
Accelerator DE-FOA-0002253, and the National Research Foundation of Korea (NRF) grant funded by the Korea government (MSIT) (RS-2022-NR068814).

\section*{Data availability}
\label{Data vailability}

The data supporting this study are available upon request.

\bibliographystyle{IEEEtran}  
\bibliography{references}     
\end{document}